\documentclass[aps,superscriptaddress,eqsecnum,nofootinbib,showpacs,preprintnumbers,twocolumn]{revtex4}
\pdfoutput=1

\usepackage{longtable}
\usepackage{bm}
\usepackage{relsize}
\usepackage{amsfonts}
\usepackage{amsmath}
\usepackage{amssymb,epsf}
\usepackage{latexsym}
\usepackage{graphicx,epsfig}
\usepackage{amssymb}
\usepackage{float}
\usepackage{subfigure}
\usepackage{epstopdf}
\usepackage[colorlinks=true,citecolor=blue,linkcolor=blue,urlcolor=black]{hyperref}
\usepackage{dcolumn}
\usepackage{psfrag}
\usepackage{wrapfig}
\usepackage{makeidx}
\usepackage{epsf}
\usepackage{color}
\usepackage{multirow}
\usepackage{mathtools}

\begin{document}

\title{Polarimetry imprints of exotic compact objects: solitonic boson stars}

\author{João Luís Rosa}
\email{joaoluis92@gmail.com}
\affiliation{Institute of Physics, University of Tartu, W. Ostwaldi 1, 50411 Tartu, Estonia}
\author{Nicolas Aimar}
\email{ndaimar@fe.up.pt}
\affiliation{Faculdade de Engenharia, Universidade do Porto, s/n, R. Dr. Roberto Frias, 4200-465 Porto, Portugal}
\affiliation{CENTRA, Departamento de Física, Instituto Superior Técnico-IST, Universidade de Lisboa-UL, Avenida Rovisco Pais 1, 1049-001 Lisboa, Portugal}
\author{Hanna Liis Tamm}
\email{hanna.liis.tamm@ut.ee}
\affiliation{Institute of Physics, University of Tartu, W. Ostwaldi 1, 50411 Tartu, Estonia}

\date{\today}

\begin{abstract} 
In this work we analyze the polarization observational properties of solitonic boson stars orbited by spherical hot-spots emitting synchrotron radiation from a thermal distribution of electrons. We consider three boson star configurations with different compacticities ranging from a dilute model with a large radius to an ultra-compact model capable of holding null bound orbits. We observe that the polarimetric imprints of the primary images for all models are comparable to those of the Schwarzschild spacetime, and thus any potentially distinguishable differences must arise from the additional structure of secondary and plunge-through images. For low inclination ($20^\circ$) observations we find that the two QU-loops of the least compact model are in contradiction with the current ALMA and GRAVITY observations and effectively excludes the possibility of Sgr A* being a dilute solitonic boson star. For high inclination ($80^\circ$) observations, both the Electric Vector Position Angle (EVPA) and the QU-loops present large qualitative deviations from the Schwarzschild black-hole for all models analyzed. Our results emphasize the suitability of polarimetry as a framework to test the nature of supermassive compact objects with future observations, especially at high inclination.
\end{abstract}

\pacs{04.50.Kd,04.20.Cv,}

\maketitle

\section{Introduction}\label{sec:intro}

Recent experimental observations in the field of gravitational physics have provided compelling evidence for the existence of ultra-compact objects in our universe. Noteworthy are the detection of gravitational wave signals from the coallescence of compact object binary systems by the Ligo-Virgo-Kagra (LVK) collaboration \cite{LIGOScientific:2016aoc,LIGOScientific:2021djp,KAGRA:2021vkt}, the observation of shadow-like dimming in the core of the galaxies M87 \cite{EventHorizonTelescope:2019dse,EventHorizonTelescope:2021bee} and the Milky Way, near Sgr A* \cite{EventHorizonTelescope:2022wkp} by the Event Horizon Telescope (EHT) collaboration, and also the observation of infrared flares orbiting close to our galactic centre by the GRAVITY collaboration \cite{GRAVITY:2020lpa,GRAVITY:2023avo}. These observations are in close agreement with the theoretical predictions in black hole (BH) spacetimes, more precisely the Kerr hypothesis \cite{Will:2014kxa,Yagi:2016jml}, which described the end-state of a full gravitational collapse in any suitable astrophysical setting as a rotating and electrically-neutral BH \cite{Kerr:1963ud,Penrose:1964wq}.

Even though BH spacetimes can successfully explain the observations outlined above, these spacetimes are intrinsically problematic from both mathematical and physical perspectives. Indeed, BH spacetimes are singular \cite{Penrose:1964wq,Penrose:1969pc}, which could imply an incompleteness of the model \cite{Romero:2013ag}. Furthermore, the event horizon incurs in the violation of the unitary evolution required by quantum mechanics \cite{Hawking:1976ra}. To address these issues, several alternative models known as exotic compact objects (ECOs) have been developed (we refer to Ref. \cite{Cardoso:2019rvt} for a complete review). Some of these ECO models are capable of reproducing similar observational predictions, thus being called BH mimickers.

One of the most popular types of ECO models are self-gravitating condensates of fundamental fields, known as boson stars \cite{Visinelli:2021uve}. These models are particularly important in comparison with other families of ECOs due to the fact that well-known dynamical formation mechanisms are known \cite{Liebling:2012fv,Brito:2015yga,Brito:2015yfh,Khlopov:1985fch}. The term "boson star" is versatile and it  encompasses a wide variety of families of models composed of different fundamental fields with different interaction potentials \cite{Brito:2015pxa,Cardoso:2021ehg}, each with their own observational implications in, e.g. X-ray spectroscopy \cite{Cao:2016zbh,Dove:1997ei}, dark matter models \cite{Hui:2016ltb}, and gravitational waves \cite{Palenzuela:2017kcg}. In particular, the absence of an event horizon is responsible for interesting effects in the gravitational wave signal, e.g. gravitational echoes \cite{Cardoso:2016oxy,Cardoso:2017cqb,Cardoso:2016rao} and tidal effects \cite{Postnikov:2010yn,Cardoso:2017cfl}. These characteristics emphasize the physical relevance of these models and justify the current active effort to analyze their observational properties \cite{Herdeiro:2021lwl,Rosa:2022tfv,Rosa:2022toh,Rosa:2023qcv,Olivares:2018abq,Gjorgjieski:2023qpv,Rosa:2024eva}. 

Recently, both the EHT~\cite{EHT_2021,EHT_2024} and the GRAVITY~\cite{gravity2018, gravity2023} experiments have successfully measured the polarization of light, i.e., the orientation on the sky of the electric field vector component of the electromagnetic wave, emitted from the vicinity of supermassive compact objects. The EHT measured this polarization directly in terms of the Stokes parameters I, Q, U, V, for the light emitted in the accretion flows of both M87* and Sgr A*, which allowed for the mapping of the magnetic field orientation and degree of order~\cite{EHT2024b}. These observations improved our understanding of astrophysics around compact objects and allowed for a better constraining of the available models. On the other hand, the GRAVITY instrument detected eight flares and measured their linear polarization~\cite{gravity2023}. Similarly to the radio data captured by ALMA~\cite{Wielgus2022}, this data shows loops of polarization in the Q-U plane, which are generated by a hot spot orbiting Sgr A*. The asymmetry in these loops is caused by general relativistic effects, with light bending having a dominant effect~\cite{Vincent:2023sbw}. The study of polarization in photon rings~\cite{Himwich2020, Palumbo2023} also indicates that polarization is sensitive to the space-time curvature. These results indicate that the analysis of light polarization serves as an adequate framework to probe the nature of compact objects. 

The comparison of the optical observational predictions of models of compact objects with photometric data requires taking into account several relativistic effects, e.g. beaming, relativistic Doppler shifting, the bending of light, among others. While analytical expressions for these effects are attainable in the Kerr geometry, the same is not true for more complicated ECO models like boson stars, for which the spacetime metric is numerical. As such, it is usual to recur to backwards ray-tracing methods to extract these effects, through which one integrates the geodesic equation for the photon from the observer to the source. Several ray-tracing codes have been developed for this purpose. In this work, we recur to the public code GYOTO~\cite{Vincent:2011wz,Aimar:2023vcs}, due to its capability of performing ray-tracing with polarization and versatility that allows the implementation of arbitrary spacetime metrics.

This work is organized as follows. In Sec. \ref{sec:theory} we introduce the Einstein-Klein-Gordon theory and solitonic boson stars as solutions in this theory; in Sec. \ref{sec:polar} we introduce the theory behind the polarimetric observables and present our results for time-integrated images and the time-evolution of polarization; and in Sec. \ref{sec:concl} we trace our conclusions.

\section{Theory and framework}\label{sec:theory}

In this work we are interested in analyzing the polarimetric signatures of scalar boson stars. These compact objects arise as solutions of self-gravitating scalar fields in the Einstein-Klein-Gordon theory, described by the following action:
\begin{equation}\label{eq:action}
    S=\int_\Omega \sqrt{-g}\left[\frac{R}{16\pi}-\frac{1}{2}\partial_\mu\Phi\partial^\mu\Phi-\frac{1}{2}V\left(|\Phi|^2\right)\right]d^4x,
\end{equation}
where $g$ is the determinant of the metric $g_{\mu\nu}$ written in terms of a coordinate system $x^\mu$, $R$ is the Ricci scalar, $\Phi$ is a complex scalar field, and $V$ is the potential of the scalar field. In Eq. \eqref{eq:action} and in what follows, we adopt a system of geometrized units such that $G=c=1$, where $G$ is the gravitational constant and $c$ is the speed of light. The field equations that describe the Einstein-Klein-Gordon system are obtained through a variation of Eq. \eqref{eq:action} with respect to the metric $g_{\mu\nu}$ and the scalar field $\Phi$, and take the forms
\begin{equation}\label{eq:field}
    G_{\mu\nu}=8\pi T_{\mu\nu},
\end{equation}
\begin{equation}\label{eq:eomphi}
    \left(\Box-\frac{dV}{d|\Phi|^2}\right)\Phi=0,
\end{equation}
where $G_{\mu\nu}\equiv R_{\mu\nu}-\frac{1}{2}g_{\mu\nu}R$ is the Einstein's tensor, $\Box\equiv \nabla_\mu\nabla^\nu$ is the d'Alembert operator, where $\nabla_\mu$ denotes the covariant derivatives, and $T_{\mu\nu}$ is the stress-energy tensor of the scalar field $\Phi$ which takes the form
\begin{equation}\label{eq:deftab}
    T_{\mu\nu}=\nabla_{(\mu}\Phi^*\nabla_{\nu)}\Phi-\frac{1}{2}g_{\mu\nu}\left(\nabla_\alpha\Phi^*\nabla^\alpha\Phi+V\right),
\end{equation}
where we have introduced index symmetrization as $X_{(\mu\nu)}=\frac{1}{2}\left(X_{\mu\nu}+X_{\nu\mu}\right)$, and $*$ denotes complex conjugation. 

In this work we are interested in static and spherically symmetric boson star solutions. As such, we introduce the following \textit{ansatze} for the metric $g_{\mu\nu}$ and the scalar field $\Phi$ in the usual spherical coordinates $x^\mu=\left(t,r,\theta,\varphi\right)$
\begin{equation}\label{eq:metric}
    ds^2=-A\left(r\right)dt^2+\frac{1}{B\left(r\right)}dr^2 + r^2d\Omega^2,
\end{equation}
\begin{equation}\label{eq:scalarfield}
    \Phi=\phi\left(r\right)e^{-i\omega t},
\end{equation}
where the metric functions $A\left(r\right)$ and $B\left(r\right)$, as well as the radial wavefunction of the scalar field $\phi\left(r\right)$ depend solely on $r$ as to preserve spherical symmetry, $d\Omega^2=d\theta^2+\sin^2\theta d\varphi^2$ represents the line-element on the two-sphere, and $\omega$ represents the angular frequency of the scalar field. Inserting Eqs. \eqref{eq:metric} and \eqref{eq:scalarfield} into Eqs. \eqref{eq:field} and \eqref{eq:eomphi}, one obtains a set of three coupled differential equations of the form
\begin{equation}\label{eq:eom1}
    \frac{B'}{r}+\frac{B-1}{r^2}=-2\pi\left(\frac{\omega^2\phi^2}{A}+B\phi'^2+V\right),
\end{equation}
\begin{equation}\label{eq:eom2}
    \frac{B A'}{r A}+\frac{B-1}{r^2}=2\pi\left(\frac{\omega^2\phi^2}{A}+B\phi'\right),
\end{equation}
\begin{equation}\label{eq:eom3}
    \frac{1}{2}\phi'\left[B\left(\frac{A'}{A}+\frac{4}{r}\right)+B'\right]+\phi\left(\frac{\omega^2}{A}-\frac{dV}{d|\Phi|^2}\right)+B\phi''=0,
\end{equation}
where a prime $(')$ denotes a derivative with respect to $r$. This system of equations is highly non-linear and thus one needs to recur to suitable numerical methods to find solutions. For this purpose, we introduce asymptotic boundary conditions that preserve the locality of the solutions, i.e., we impose the asymptotic flatness of the spacetime, and the vanishing of the radial wavefunction of the scalar field at infinity, i.e., 
\begin{eqnarray}\label{eq:bcAinf}
    A\left(r\to\infty\right)&=&1-\frac{2M}{r},\\ \label{eq:bcBinf}
    B\left(r\to\infty\right)&=&1-\frac{2M}{r},\\ \label{eq:bcphiinf}
    \phi\left(r\to\infty\right)&=&0, 
\end{eqnarray}
where $M$ is the total mass of the boson star. On the other hand, to preserve the regularity of the solutions at the origin, we impose the following set of boundary conditions at the origin
\begin{eqnarray}\label{eq:bcA0}
    A\left(r\simeq 0\right)&=&A_0,\\ \label{eq:bcB0}
    B\left(r\simeq 0\right)&=&A_0,\\ \label{eq:bcphi0}
    \phi\left(r\simeq 0\right)&=&\phi_0, \qquad \phi'\left(r\simeq 0\right)=0.
\end{eqnarray}
We note that the constant $A_0$ can always be set to zero through a time reparametrization, resulting in a solution that does not satisfy the boundary condition in Eq. \eqref{eq:bcAinf}. Nevertheless, after the solution has been found, the time coordinate can be rescaled through a modification of $A$ and $\omega$ as to satisfy the asymptotic boundary condition. 

In order to integrate the system of equations in Eqs. \eqref{eq:eom1} to \eqref{eq:eom3} under the boundary conditions given in Eq. \eqref{eq:bcAinf} to \eqref{eq:bcphi0}, it is necessary to specify a form for the potential $V$. In previous works \cite{Macedo:2013jja,Rosa:2023qcv,Rosa:2024eva}, it was shown that the so-called solitonic potential is particularly useful to obtain ultra-compact boson star solutions, known as solitonic boson stars, that resemble the black-hole spacetime the most. Thus, in this work, we focus our attention towards solitonic boson stars, which are described by the potential \cite{Lee:1986ts} 
\begin{equation}\label{eq:potential}
    V\left(|\Phi|^2\right)=\mu^2|\Phi|^2\left(1+\frac{|\Phi|^2}{\alpha^2}\right)^2,
\end{equation}
where $\mu$ is a constant that plays the role of the mass of $\Phi$ and $\alpha$ is a constant free parameter of the model. The most compact solutions for this model are obtained in the limit $\alpha\to 0$, with a minimum radius of $R\simeq 2.81M$.\cite{Lee:1991ax,Cardoso:2021ehg}, where the radius $R$ is defined as the radius encapsulating $98\%$ of the mass of the boson star. For the purpose of this work, we select three configurations with $\alpha=0.08$ that have been previously used in other works. These solutions are summarized in Table \ref{tab:solutions} (see also Refs. \cite{Rosa:2023qcv,Rosa:2024eva} for additional details). 

\begin{table}[]
    \centering
    \begin{tabular}{c c c c c c}\hline
        Configuration & $\phi_0$ & $\mu M$  & $\mu R$ & $\mathcal C$ & $\omega/\mu$ \\ \hline
        SBS1 & 0.0827 & 1.7531 & 11.5430 & 0.1518 & 0.25827 \\
        SBS2 & 0.0827 & 4.220 & 16.6520 & 0.25342 & 0.17255 \\
        SBS3 & 0.0850 & 5.655 & 17.6470 & 0.32045 & 0.13967 \\ \hline
    \end{tabular}
    \caption{Details on the boson star configurations considered. The compacticity is defined as $\mathcal C\equiv M/R$ and its maximum value for the Schwrazschild black-hole is $\mathcal C_{\rm max}=0.5$.}
    \label{tab:solutions}
\end{table}

\section{Polarimetry}\label{sec:polar}

\subsection{Polarimetric observables}
To analyze the polarimetric signatures of the boson star configurations introduced above, we recur to the ray-tracing software GYOTO \cite{Vincent:2011wz,Aimar:2023vcs}. GYOTO outputs a set of 2-dimensional matrices representing the specific intensities of the Stokes parameters. In particular, we are interested in two Stokes parameters, namely $Q$ and $U$, which are defined as follows~\cite{Vincent:2023sbw}. Consider the electric field vector of an incident wave on the observer's screen
\begin{equation}
    \textbf{E}=E\left(\cos \chi_o \textbf{e}_\alpha+\sin\chi_o \textbf{e}_\beta\right),
\end{equation}
where $E$ is the amplitude of the electric field, $\chi_o$ is the observed electric vector position angle (EVPA), and the vectors $\left(\textbf{e}_\alpha,\textbf{e}_\beta\right)$ are an orthonormal basis in the plane of the observer's screen. The Stokes parameters $Q$ and $U$ are defined as
\begin{eqnarray}
    Q&=&I\cos\left(2\chi_o\right),\\
    U&=&I\sin\left(2\chi_o\right),
\end{eqnarray}
where $I=E^2$ is the total intensity of the incident wave, also known as the Stokes I parameter. The observed EVPA is thus given by
\begin{equation}
    \chi_o=\frac{1}{2}{\rm atan2}\left(Q,U\right),
\end{equation}
which lies in the interval $\chi_o\in\left[-\frac{\pi}{2},\frac{\pi}{2}\right]$.

The polarization observed depends on two factors: 1) the emission mechanism, which corresponds in this case to synchrotron radiation; and 2) the curvature of the null geodesics along which photons propagate. In the rest frame of the emitter, the polarization vector for synchrotron radiation $\mathbf{f_e}$ is simultaneously orthogonal to the wave vector of the photon $\mathbf{K_e}$ and the magnetic field vector $\mathbf{B_e}$~\citep{RybickiLightman:1979}. This can be expressed through
\begin{equation} \label{eq:polar_vector}
    \mathbf{f_e} = \mathbf{K_e} \times \mathbf{B_e}.
\end{equation}
This polarization vector can be expressed in terms of the Stokes parameters (see Ref.~\cite{Marszewski:2021}). In vacuum, this vector is always orthogonal to the direction of propagation of the photons which, in curved spacetime, changes along the null geodesics. This vector must thus be parallel transported from the emitter to the observer. While this parallel transport can be done analytically for the Kerr metric~\cite{Gelles:2021}, the same is unachievable for numerical boson star spacetimes, and thus we recur to GYOTO to perform this task (see \cite{Aimar:2023vcs} for more details).

The Stokes parameters output by GYOTO are given in 2-dimensional matrices of specific intensities $S_{lm}^\nu$, where $S=\{I,Q,U\}$, for each time instant $t_k$. The indices $\{m,l\}$ represent the pixels of the image associated with an observed Stokes parameter. Repeating the simulation through several time instants $t_k\in\left[0,T\right[$, where $T$ is the orbital period of the source, one obtains a cube of data $S_{klm}$, where the index $k$ covers the time instants and the indices $lm$ cover the pixels on the observer's screen. For each of the cubes of data representing each of the Stokes parameters, the time integrated flux is given by
\begin{equation}
    \left<S\right>_{lm} = \sum_k S_{klm}.
\end{equation}

In what follows, the comparison between different boson stars models, as well as with the results in the Schwarzschild spacetime, is done through the analysis of the following three observables: the time-integrated Stokes parameters $\left<S\right>$, the QU-loops $U\left(Q\right)$, and the temporal EVPA $\chi_o\left(t\right)$.

\subsection{Numerical setup}
We simulate the orbit of an emitting spherical light source of radius $R_s=0.5M$ around a central object described by a solitonic boson star of ADM mass $M=4.2\times 10^6 M_\odot$ seen by an observer at $d=8.25$kpc at a frequency of 230GHz. The resolution of the images generated is 1000x1000 pixels with a field of view of 250$\mu as$. Such high resolution is needed to resolve secondary and plunge-through images, the latter being a feature of some solitonic boson star models.\\

The source orbits at the equatorial plane $\theta=\pi/2$ with a constant orbital radius $r_o=8M$ with a Keplerian velocity. The source is emitting synchrotron radiation from a thermal distribution of electrons at a dimensionless temperature of $\Theta_e=200$ and a number density of $n_e=6.6$ cm$^{-3}$. Following the results from ALMA flare observation \cite{Wielgus2022} and GRAVITY flare data \cite{GRAVITY:2023avo}, we consider (if not specified otherwise) a vertical magnetic field configuration with a strength of $B\approx 0.34$G, i.e. a magnetization parameter of $\sigma=0.01$.

\subsection{Results}

\subsubsection{Time-integrated images}

\begin{figure*}
    \centering
    \includegraphics[width=0.32\linewidth]{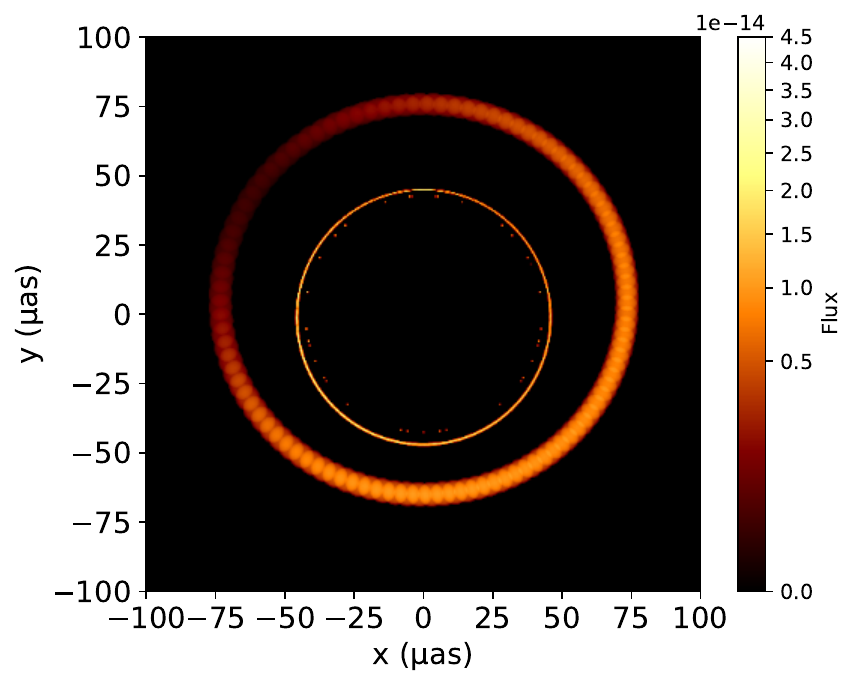}\quad
    \includegraphics[width=0.32\linewidth]{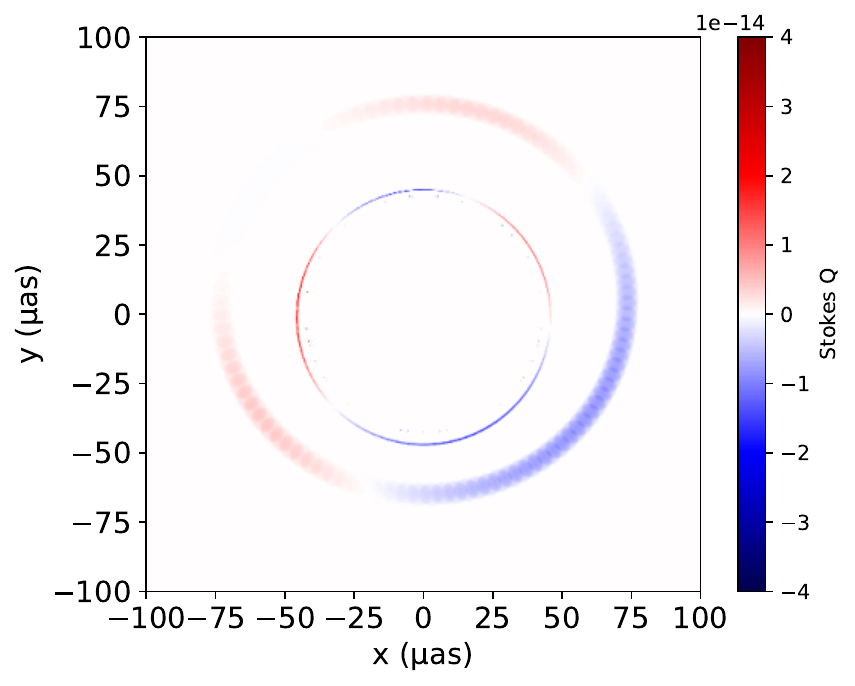}\quad
    \includegraphics[width=0.32\linewidth]{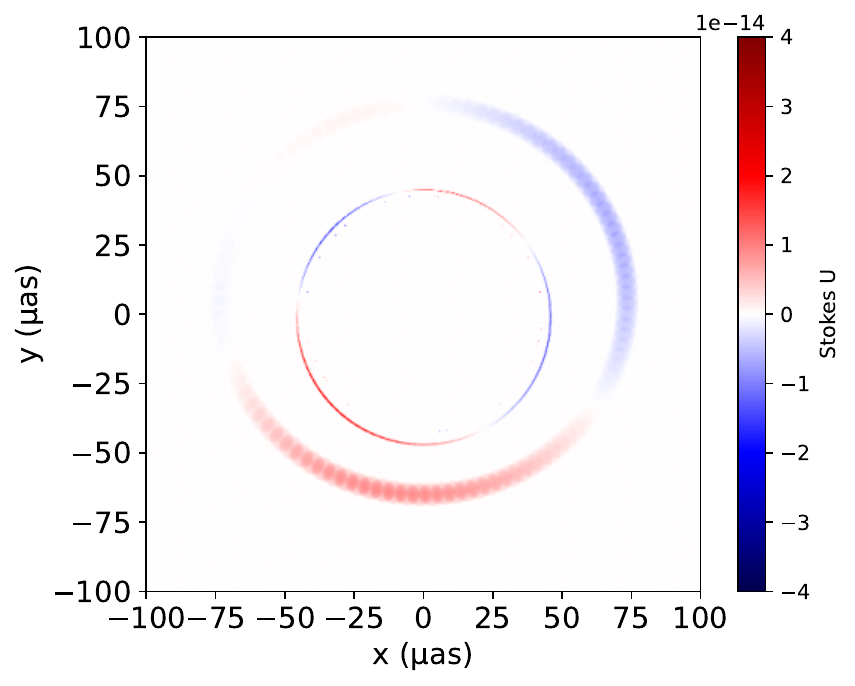}\\
    \includegraphics[width=0.32\linewidth]{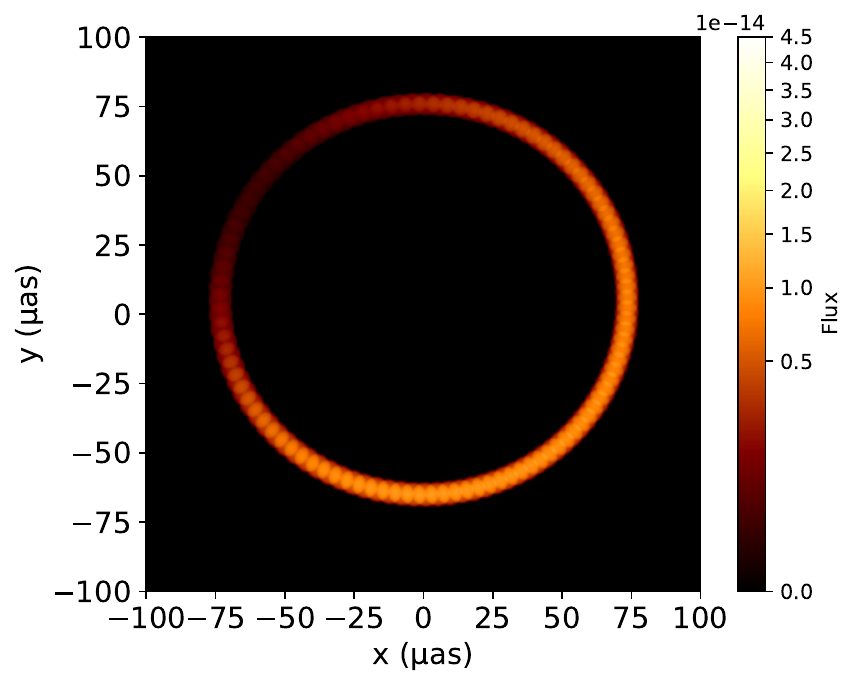}\quad
    \includegraphics[width=0.32\linewidth]{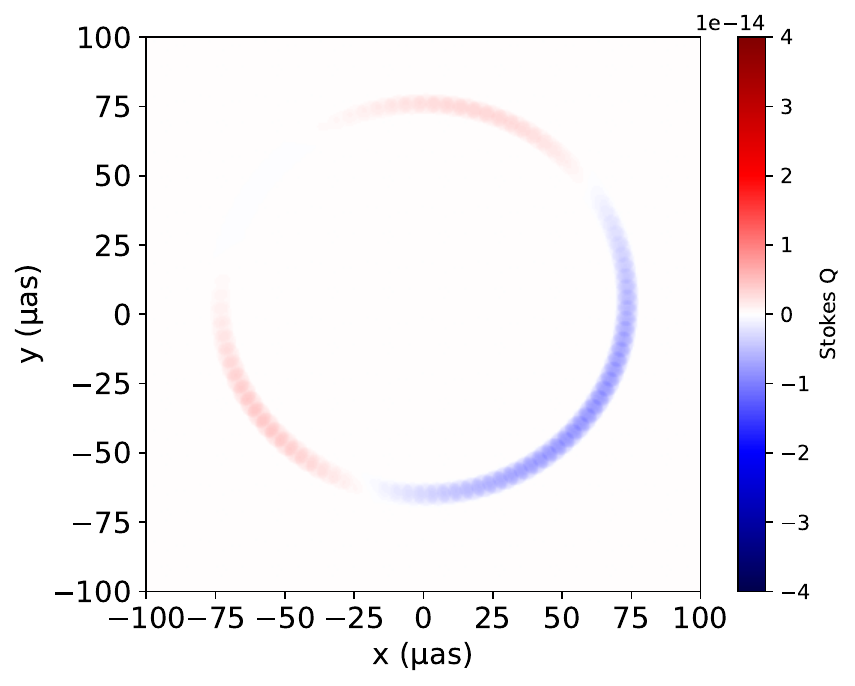}\quad
    \includegraphics[width=0.32\linewidth]{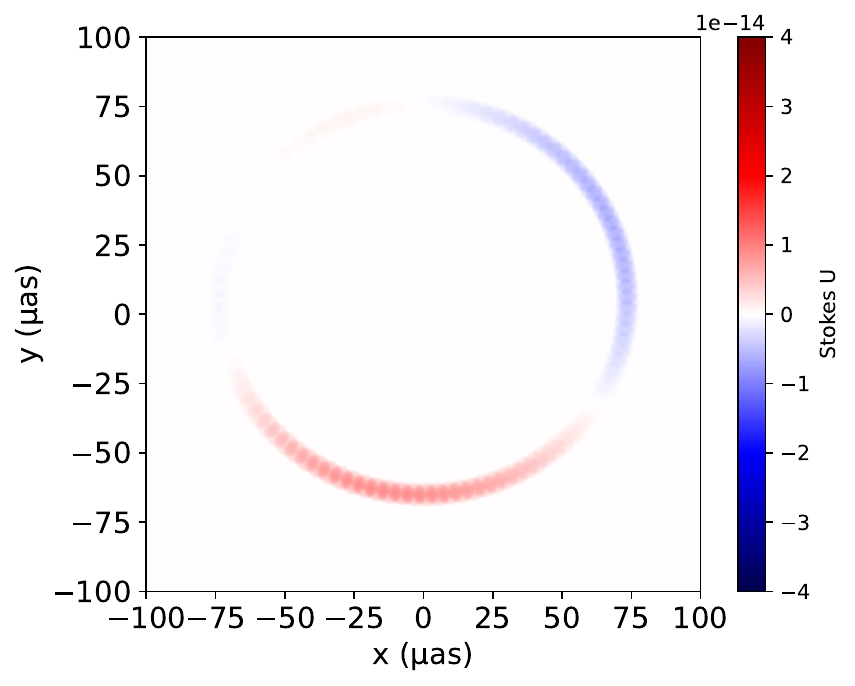}\\
    \includegraphics[width=0.32\linewidth]{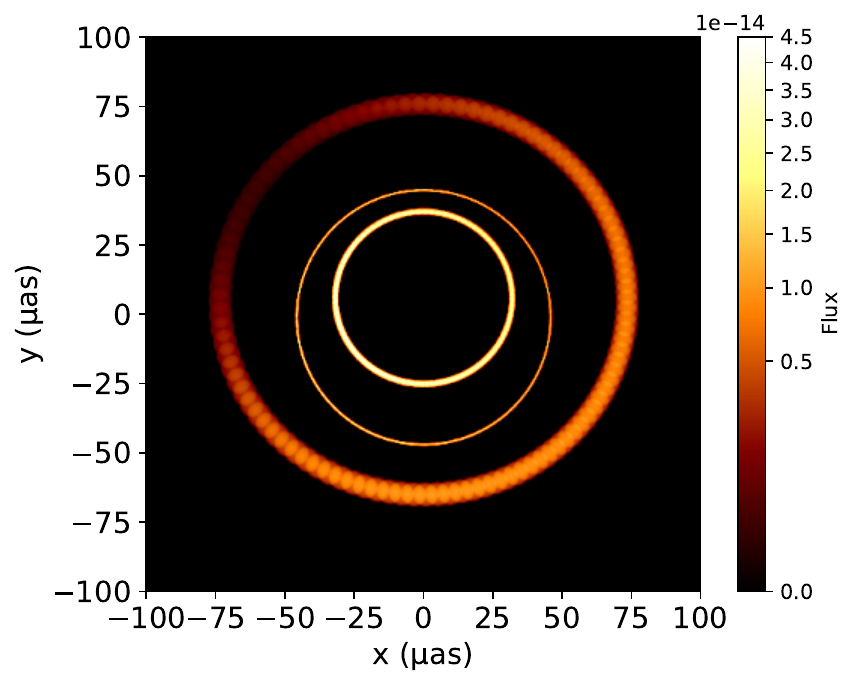}\quad
    \includegraphics[width=0.32\linewidth]{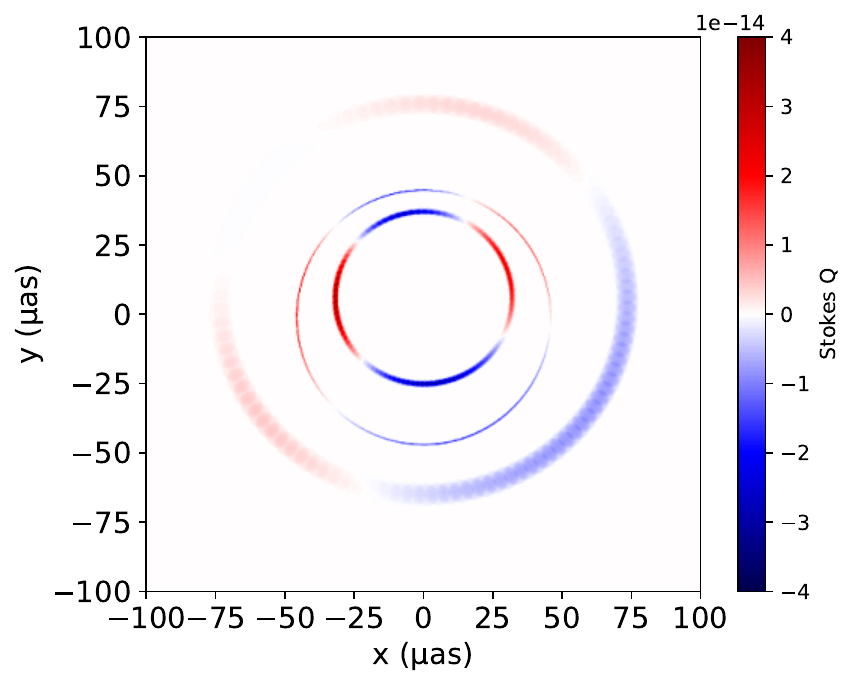}\quad
    \includegraphics[width=0.32\linewidth]{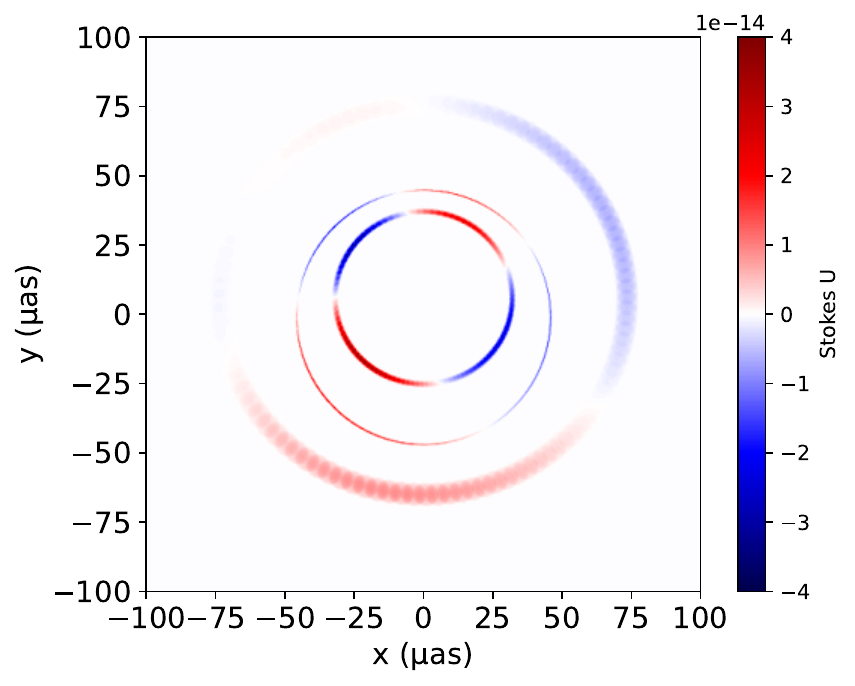}\\
    \includegraphics[width=0.32\linewidth]{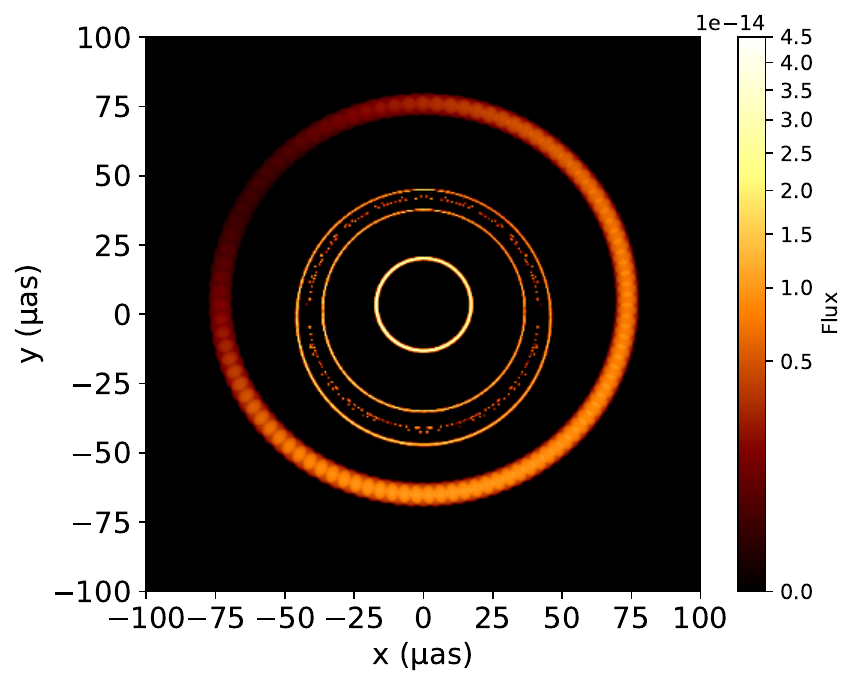}\quad
    \includegraphics[width=0.32\linewidth]{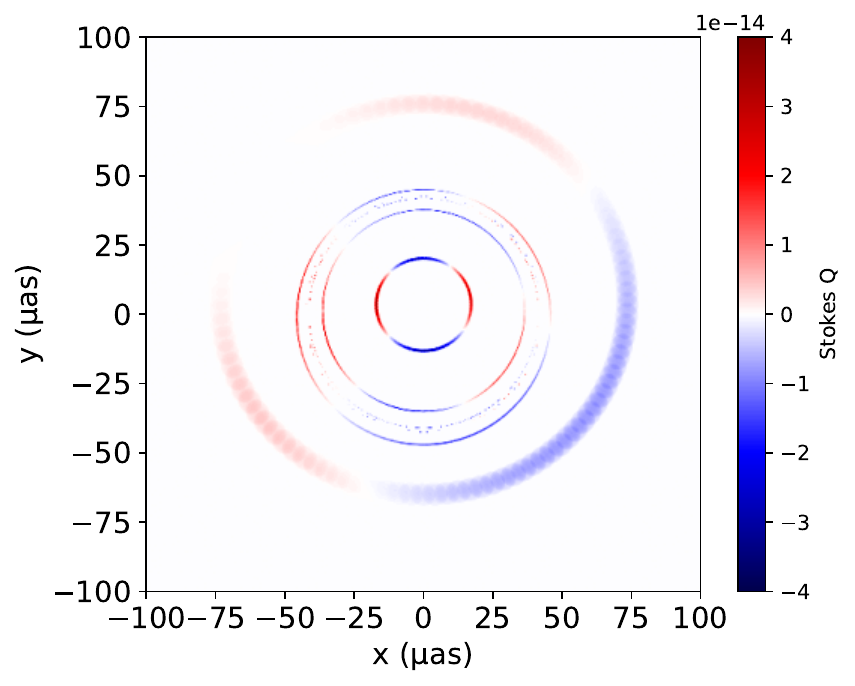}\quad
    \includegraphics[width=0.32\linewidth]{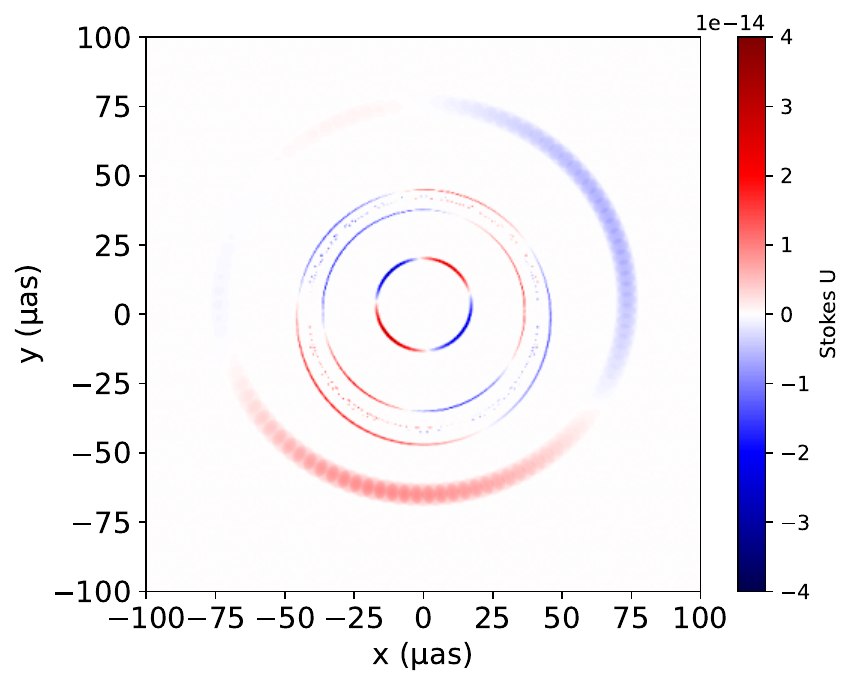}
    \caption{Integrated flux of the Stokes parameters I (left column), Q (middle column) and U (right column) for the Schwarzschild BH (first row), and for the boson star models SBS1 (second row), SBS2 (third row), and SBS3 (fourth row), for an observation inclination of $20^\circ$.}
    \label{fig:stokes_20}
\end{figure*}

\begin{figure*}
    \centering
    \includegraphics[width=0.32\linewidth]{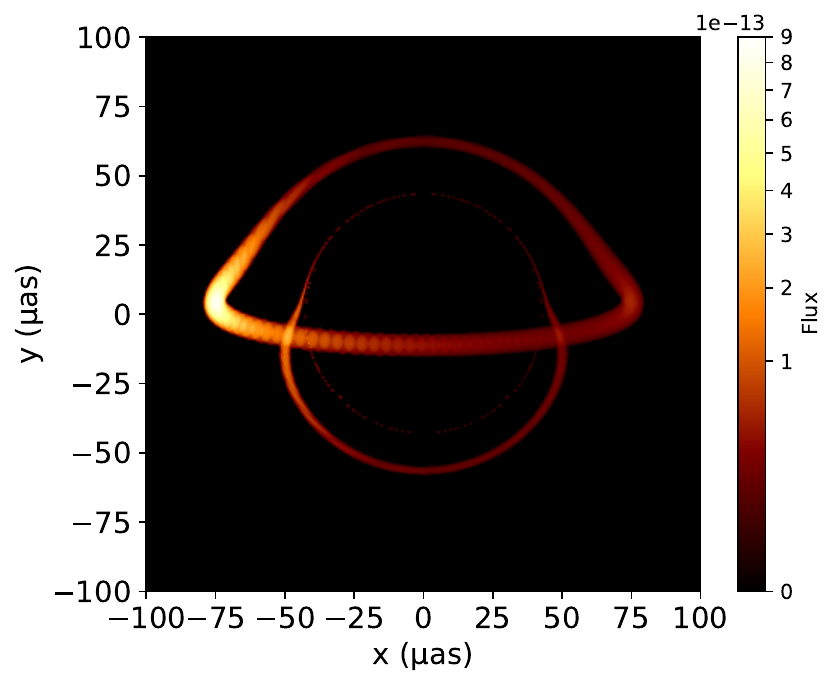}\quad
    \includegraphics[width=0.32\linewidth]{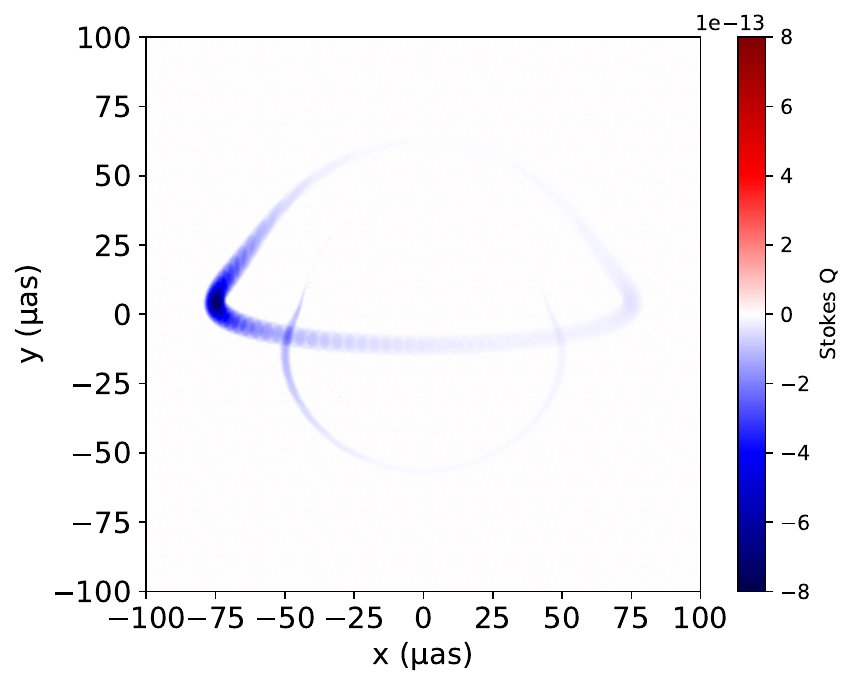}\quad
    \includegraphics[width=0.32\linewidth]{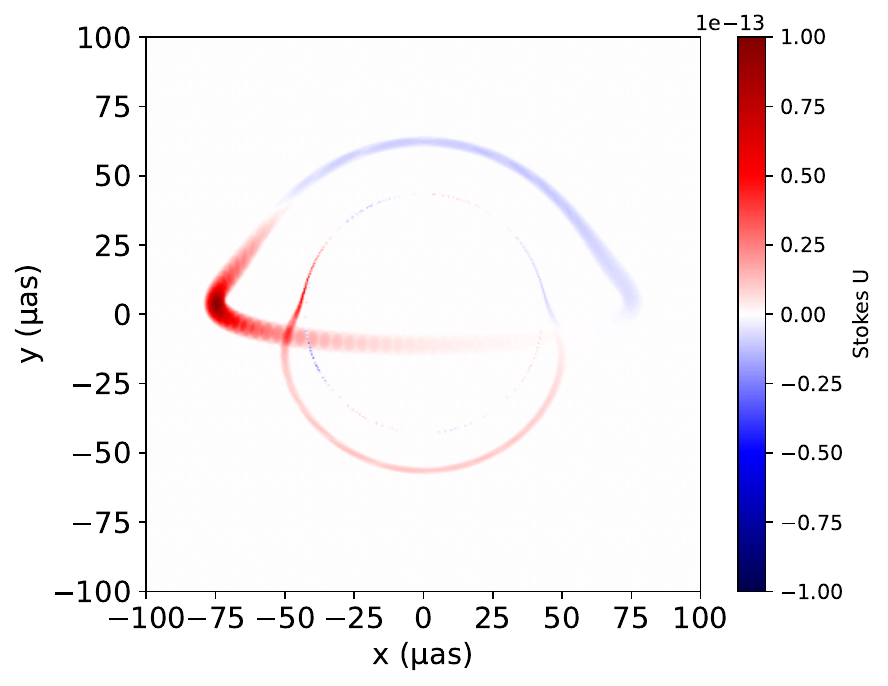}\\
    \includegraphics[width=0.32\linewidth]{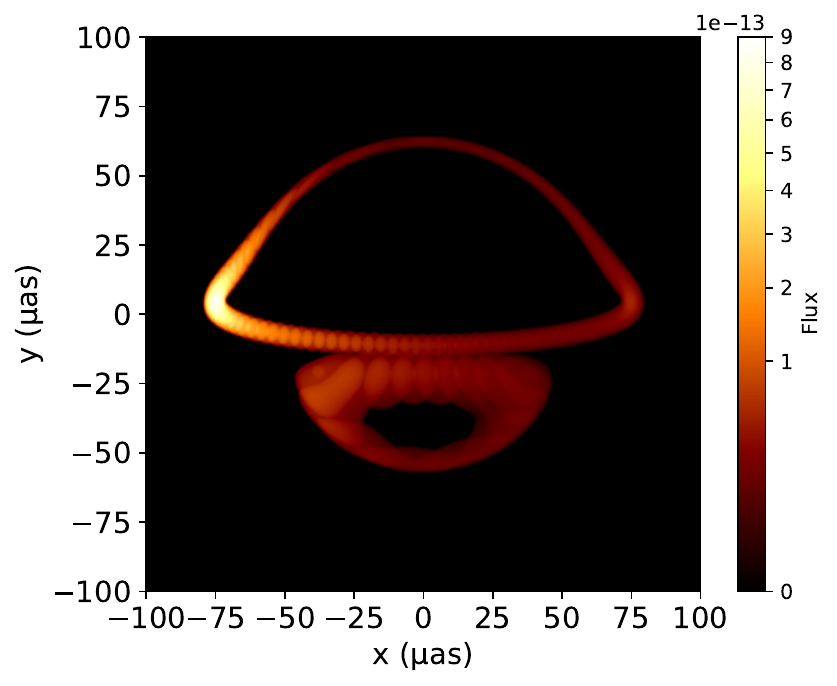}\quad
    \includegraphics[width=0.32\linewidth]{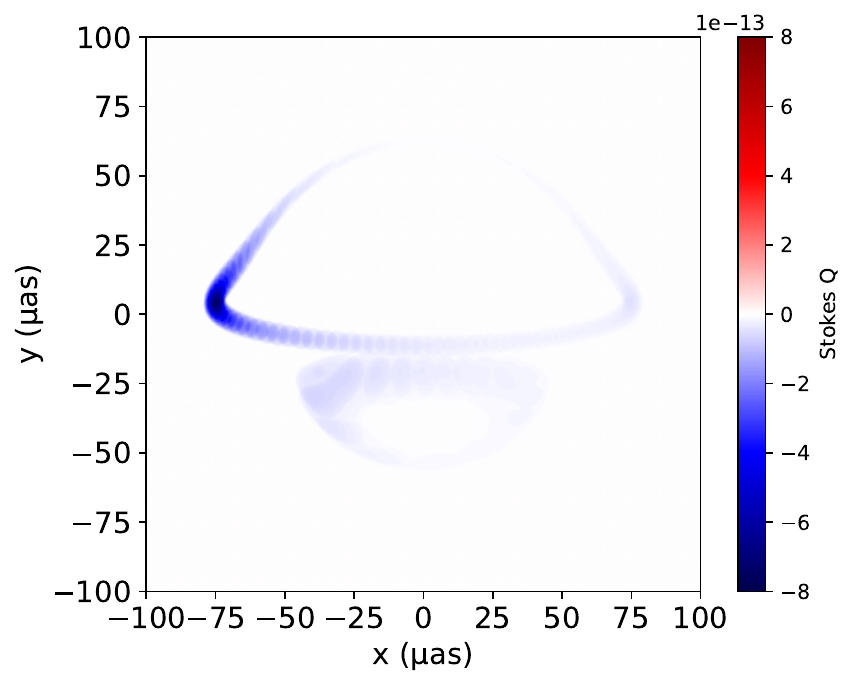}\quad
    \includegraphics[width=0.32\linewidth]{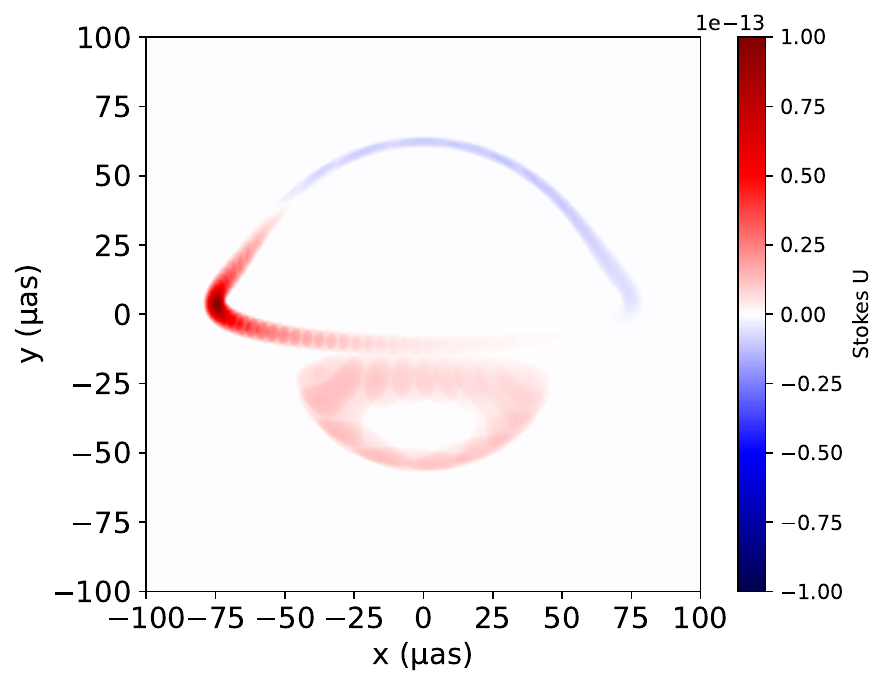}\\
    \includegraphics[width=0.32\linewidth]{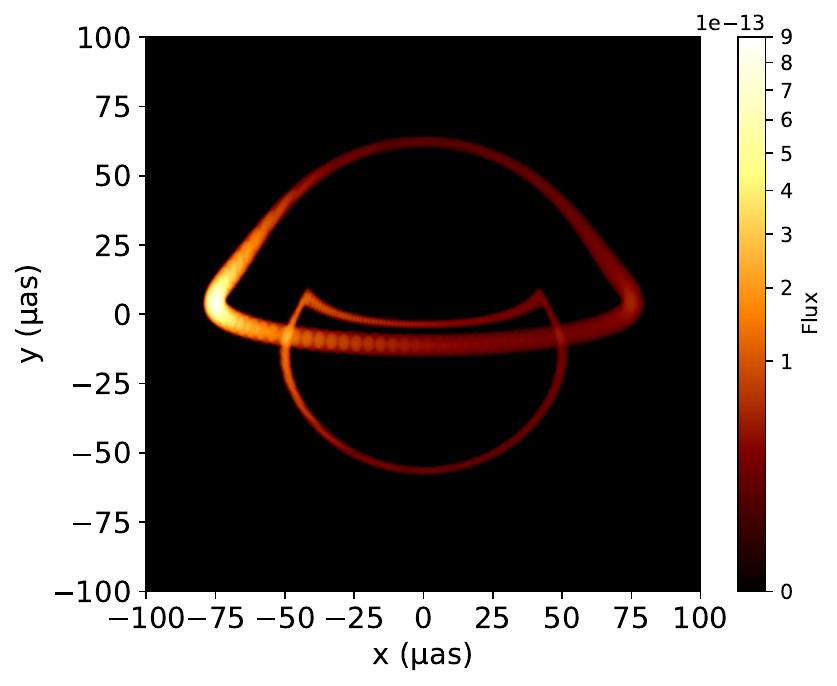}\quad
    \includegraphics[width=0.32\linewidth]{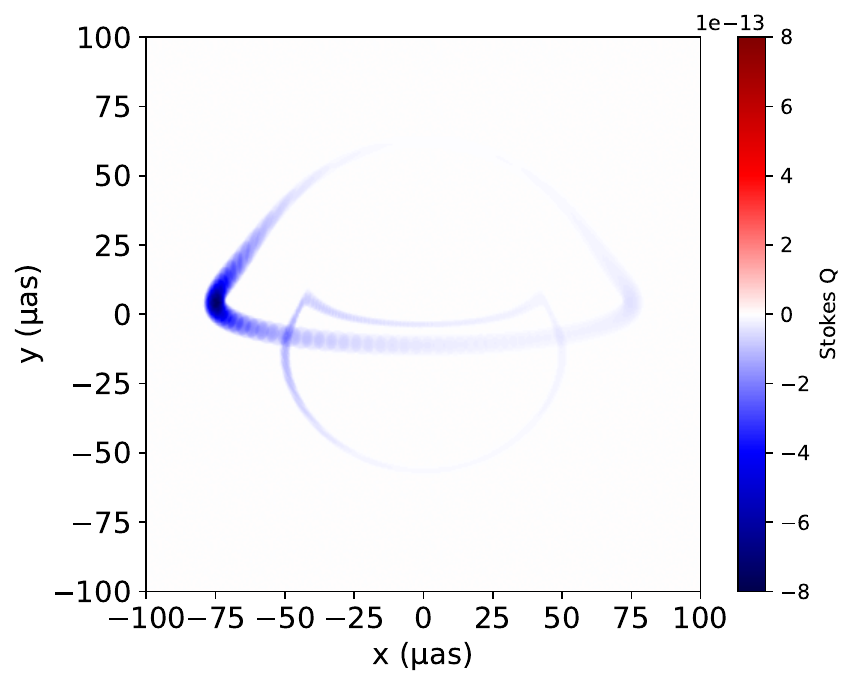}\quad
    \includegraphics[width=0.32\linewidth]{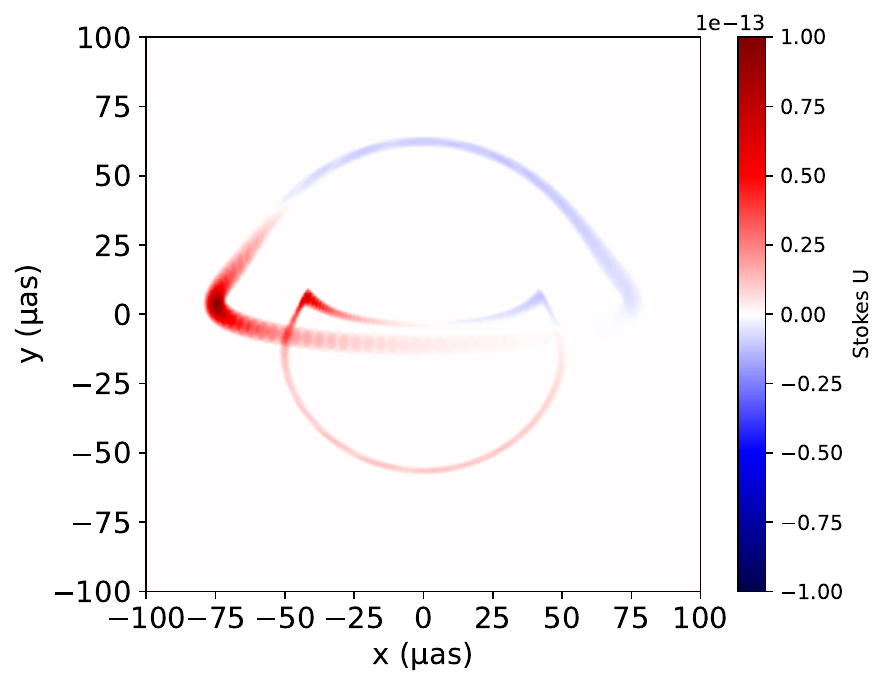}\\
    \includegraphics[width=0.32\linewidth]{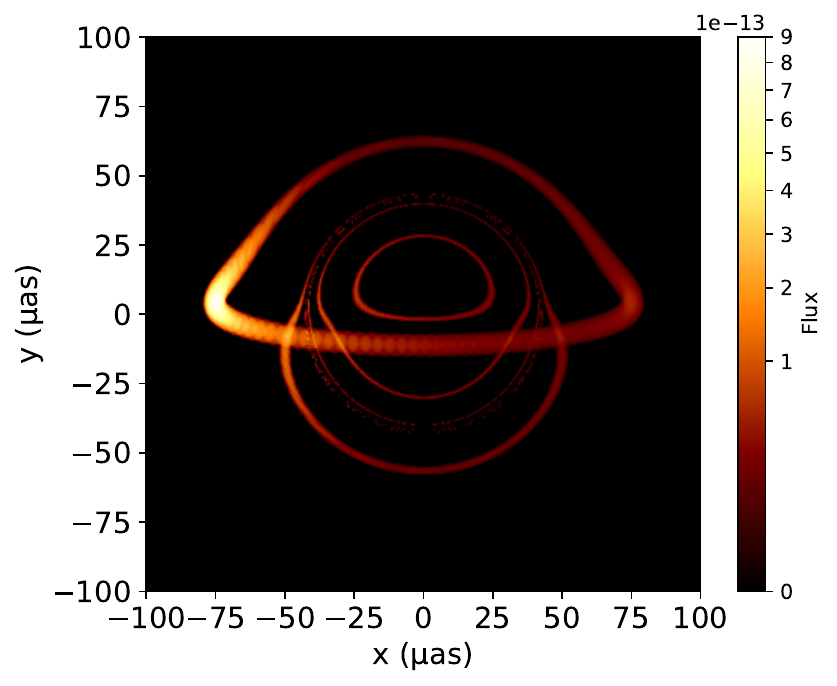}\quad
    \includegraphics[width=0.32\linewidth]{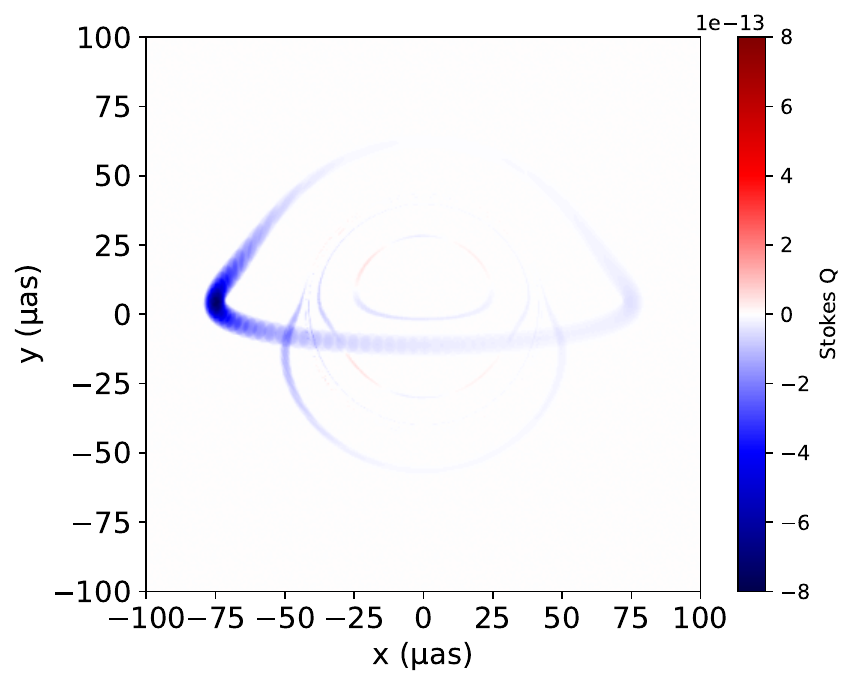}\quad
    \includegraphics[width=0.32\linewidth]{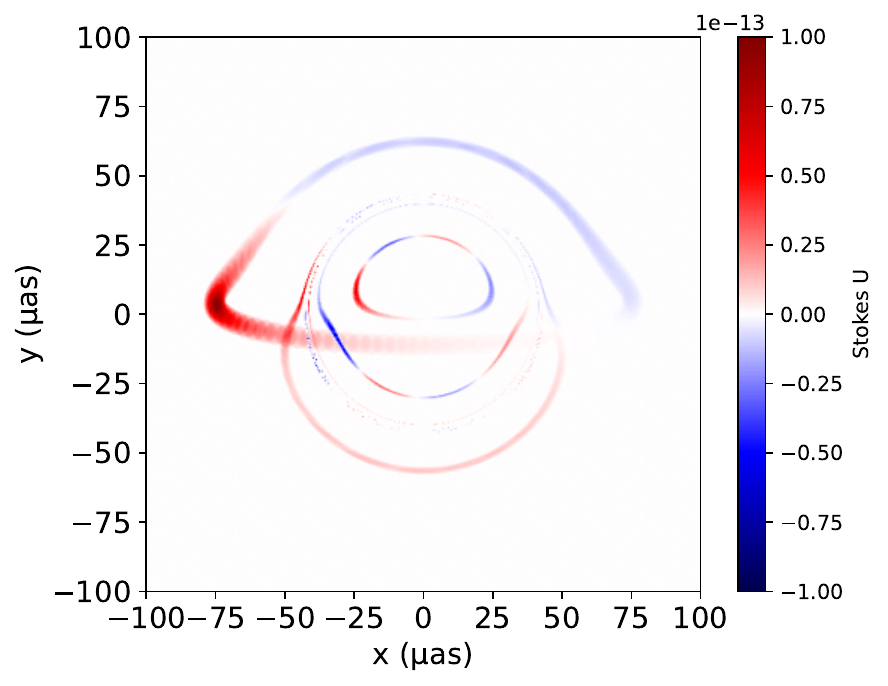}
    \caption{Integrated flux of the Stokes parameters I (left column), Q (middle column) and U (right column) for the Schwarzschild BH (first row), and for the boson star models SBS1 (second row), SBS2 (third row), and SBS3 (fourth row), for an observation inclination of $80^\circ$.}
    \label{fig:stokes_80}
\end{figure*}

The images generated by the time-integrated Stokes parameters for the Schwarzschild BH and for the three SBS models considered, for observation inclinations of $20^\circ$ and $80^\circ$, are given respectively in Figures \ref{fig:stokes_20} and \ref{fig:stokes_80}. These results show different image structures depending on the model being simulated. 

Consider the results for an inclination of $20^\circ$. For all of the models, a primary image track is visible, corresponding to the outer approximately circular contribution. In comparison with previous studies which assume isotropic emission, synchrotron radiation in a specific magnetic field configuration is not isotropic as the radiative coefficients depend on the angle $\theta_m$ between the direction of the photon and the magnetic field~\cite{RybickiLightman:1979,Marszewski:2021}. This dependence explains the dim region in the top-left part of the intensity maps (Fig~\ref{fig:stokes_20}). For this part of the images, due to light bending, $\theta_m$ is closer to zero in comparison with the rest of the images, resulting in a lower intensity. Note that this feature is present in the four modeled metrics, i.e. the Schwarzschild and the three SBS metrics.

The ultracompact configurations, i.e., the Schwarzschild BH and the configuration SBS3 feature a thin photon-ring contribution, also known as the critical curve, corresponding to the photons that are asymptotically bound to the unstable photon orbit. The Schwarzschild BH and the configurations SBS2 and SBS3 feature a secondary image track with an impact parameter larger than that of the critical curve. Due to the absence of an event horizon, the configurations SBS2 and SBS3 feature additional secondary tracks with impact parameters smaller than that of the critical curve, also known as plunge-through images, with SBS2 featuring one additional secondary track and the SBS3 featuring two additional secondary tracks. 

It is noteworthy that the photons from primary images, as well as from the exterior secondary images and the photon-ring contributions, whenever the latter are present, show the same polarization characteristics with alternating polarity which depends on the position on the sky and the order of the image. In particular, at low inclination, the secondary and plunge through images have the similar polarization for the same position angle on-sky. This indicates that it is the presence or absence of additional images that induces the most prominent differences in the observed polarization signal. We provide evidence to support this statement in what follows, when we introduce the time-evolution of the polarization signal, by comparing the polarimetric observables in the presence and in the absence of higher-order images. 

Regarding the results for an inclination of $80^\circ$, there are two noteworthy features that differ from the $20^\circ$ case. First, the secondary track is now visible for the SBS1 configuration. This happens because, since the observation inclination is larger, the amount of light deflection necessary for the appearance of a secondary image is smaller, and thus even the least compact models are capable of producing this feature (see Fig. 6 in Ref. \cite{Rosa:2023qcv}). Second, the two secondary tracks in the SBS2 configuration merge into a single track. This is an indication that the secondary images are not always visible at this inclination, and they appear only while the source is moving on the opposite side to the observer, i.e., behind the compact object. For the Schwarzschild BH and the SBS3 configurations, the number and presence of the image tracks are the same compared to the $20^\circ$ inclination case, although distorted due to the effects of light deflection.

\subsubsection{Time evolution of the polarization}
\begin{figure*}
    \centering
    \includegraphics[width=0.4\linewidth]{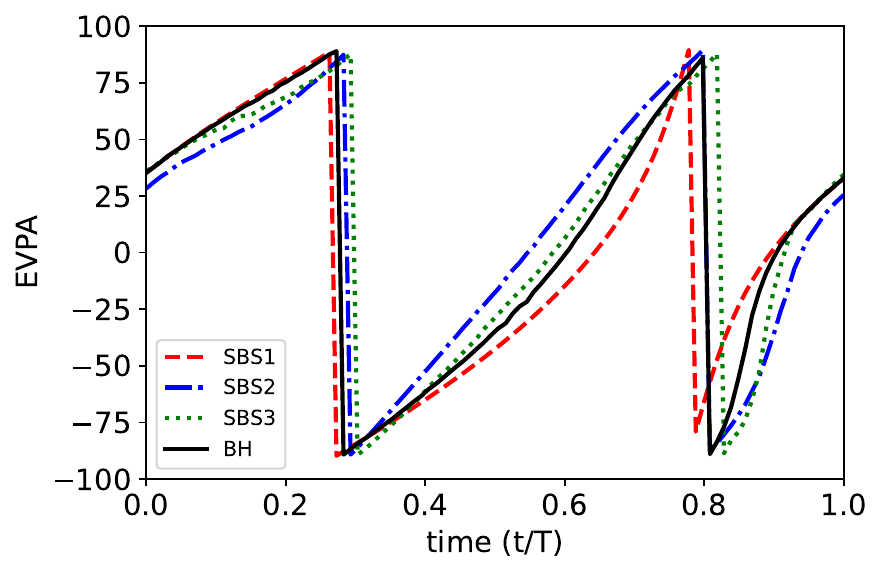}\qquad
    \includegraphics[width=0.4\linewidth]{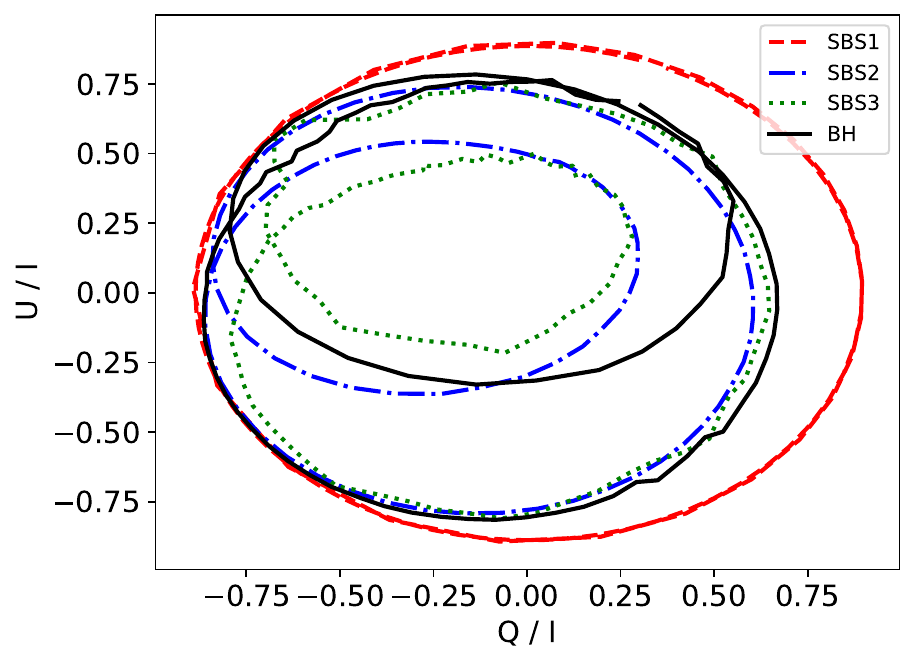}\\
    \includegraphics[width=0.4\linewidth]{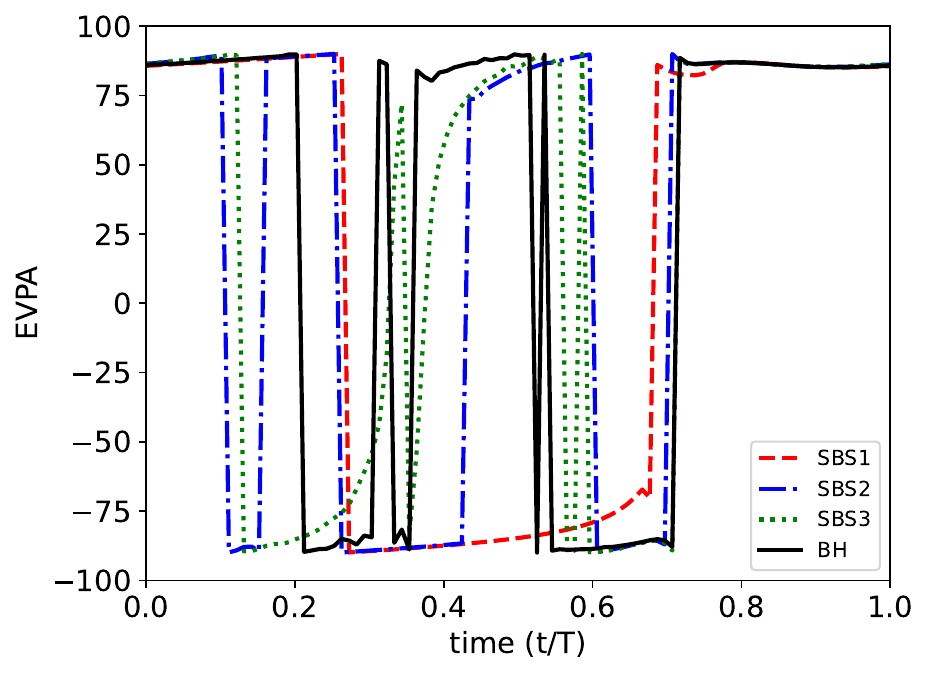}\qquad
    \includegraphics[width=0.4\linewidth]{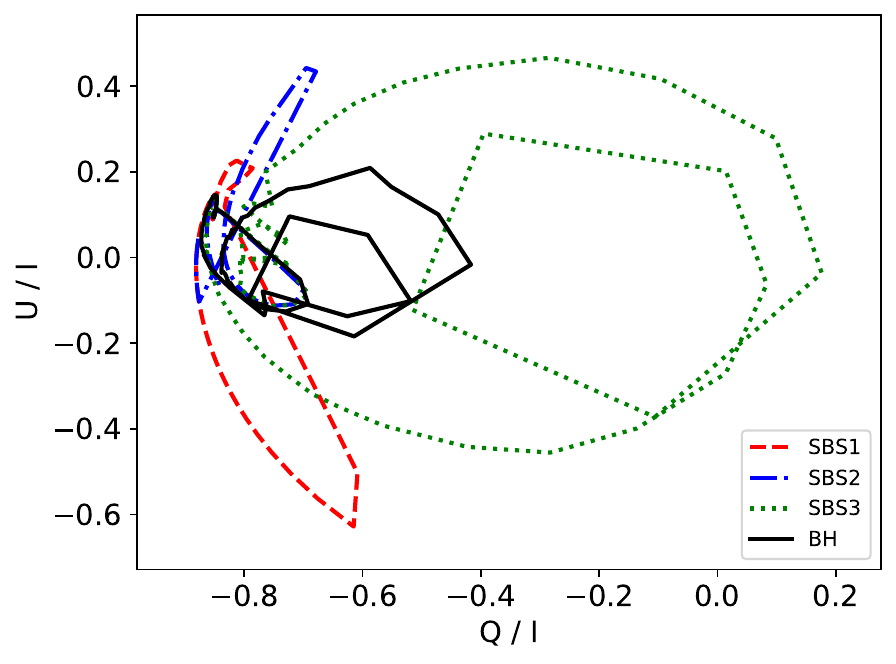}
    \caption{EVPA as a function of the normalized orbital time (left column) and loops in the QU-plane (right column) for an observation inclination of $20^\circ$ (top row) and $80^\circ$ (bottom row).}
    \label{fig:EVPA_QU}
\end{figure*}

We focus now on the temporal properties of the observed (image integrated) polarization. We show in Figure \ref{fig:EVPA_QU} the observed temporal EVPA and QU-loops for the inclinations of $20^\circ$ and $80^\circ$ in the four modeled metrics.

\begin{figure*}
    \centering
    \includegraphics[width=0.8\linewidth]{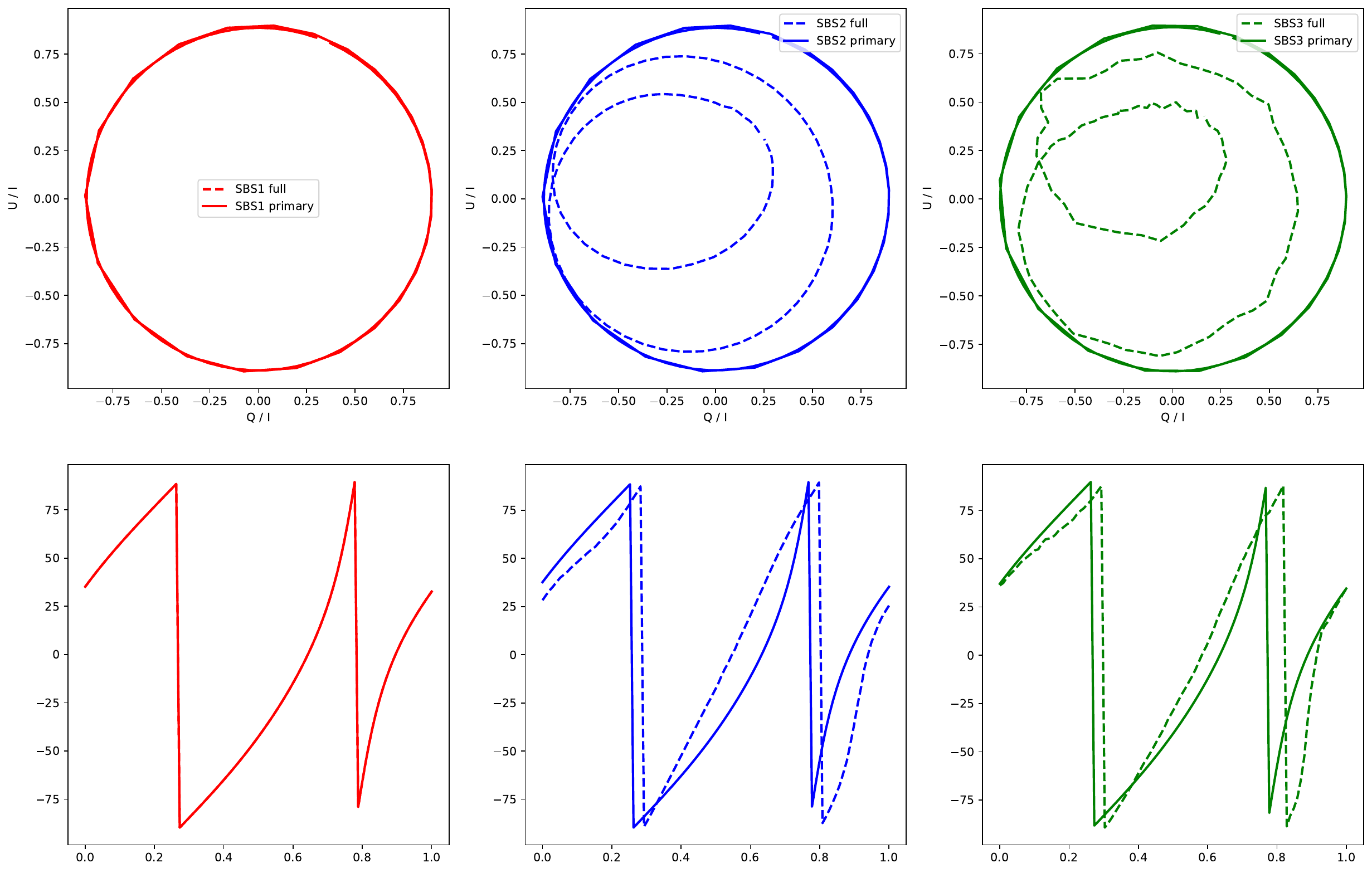}
    \caption{Loops in the QU-plane (top row) and EVPA as a function of the normalized orbital time (bottom row) for the SBS1 (left column), SBS2 (middle column) and SBS3 (right column) models, for an observation inclination of $20^\circ$, when only the primary image is considered (solid lines) and when the primary and higher-order images are considered simultaneously (dashed lines).}
    \label{fig:QUcomparison}
\end{figure*}

For the $20^\circ$ inclination and for all metrics, the overall EVPA curves as functions of time show a decrease at an almost constant rate with two inversions over one orbital period. The EVPA values are in [$-\frac{\pi}{2}$, $\frac{\pi}{2}$], thus these two inversions mean that the polarization vector is making a full $360^\circ$ revolution on-sky in one orbital period as observed by \cite{GRAVITY:2023avo}. However, the time between the two inversions and the shape of the EVPA curve are not the same for all the metrics. The presence or absence of secondary and plunge-through images not only impacts the time-integrated images but also the observed EVPA and QU-loops. To clarify this property, Fig. \ref{fig:QUcomparison} compares the observed EVPA and the QU-loops for each model when only the primary image is considered, and when the full image structure, including the higher-order images, is considered. Indeed, one observes that if only the primary image is considered, the observed EVPA and QU-loops are the same for every model, whereas distinguishable properties of these quantities arise when the additional higher-order images are considered.

Indeed, for the SBS1 configuration in which only the primary image is visible at $20^\circ$ of inclination, one observes that the shifting time of the EVPA is slightly shorter and the EVPA curve is lower between inversions in comparison with the other models. Furthermore, the QU-loops differ the most in comparison with the BH scenario, with both loops showing approximately the same larger shape, as happens for the other models when only the primary image is considered. The latter result is in contradiction with the ALMA and GRAVITY observations, for which the second loop has a largely smaller radii compared to the first one \cite{Wielgus2022}. This result seems to indicate that the SBS1 model is unsuitable to explain the available experimental data.

On the other hand, for the SBS2 and SBS3 configurations in which additional secondaries and plunge-through images are present, these images seem to slightly affect the time between the two shift of the EVPA. As for the QU-loops, the presence of these additional images causes the second QU-loop to be smaller than the first. Indeed, the plunge-through images can have an opposite polarization in comparison with the primary image, which on one hand causes the total unpolarized intensity to increase in comparison with the BH case, but on the other hand they can cause a decrease in the integrated value of the Stokes Q and U, reducing the polarization fraction. Because of this effect and the difference in the sizes of the plunge-through images in the SBS2 and SBS3 models, the BS3 configuration presents the smallest loop of the set of models analyzed.

Unlike it happens for low inclinations, for which the intensity of the Stokes Q and U parameters are of the same order of magnitude, for high inclinations the Stokes Q parameter is one order of magnitude larger. This is because the magnetic field is vertical, and the photon direction is almost perpendicular to the latter, with approximately the same orientation (neglecting light bending and special relativistic effects). The observed polarization vector is thus horizontal almost everywhere in the primary image (Stokes Q $<$ 0). The light bending and beaming effect are strong at high inclination, but that affects mostly the intensity and not the EVPA, which is governed by the geometry of the system (see Eq.~\ref{eq:polar_vector}). These two effects can be seen in the Stokes U image where the beaming increase the intensity when the plasma is moving in the direction of the observer, and light bending changes the sign of Stokes U depending on the position of the source, with $U>0$ when the source is between the observer and the central object and $U<0$ when the source is behind the central object.

The effects of the additional images in the EVPA and the QU-loops, for which light bending is much stronger, becomes especially prominent. Unlike it happens for low inclinations, for which all models present the same number of EVPA inversions although with different time intervals between inversions, the number of EVPA inversions for high inclination changes drastically between models. These inversions, however, are caused by the small changes of Stokes U (going from positive to negative values in vice-versa) which are strongly affected by numerical errors in most models. Due to the fact that the observables for the SBS1, SBS2, and BH models are strongly dominated by negative Q values, the trajectories on the QU-plane cross the $U=0$ line in the region where $Q<0$ several times, causing the inversions in the EVPA to happen. As such, the only model for which one can extract useful information from the EVPA is SBS3, where the two observed QU-loops can be translated into the smooth transitions in the EVPA that happen in the time intervals between $t\sim 0.1 T$ to $t\sim 0.3T$, followed by an inversion, and another smooth transition from $t\sim 0.3T$ to $t\sim 0.6T$.

Furthermore, at large inclinations, we observe that only the ultracompact SBS models, i.e., the SBS3 configuration, presents photon contributions for which the Stokes Q parameter is positive, corresponding to the inner secondary contributions. These features cause the QU-loops for the SBS3 configuration to be wider in the QU plane than those of the configurations SBS1 and SBS2. In particular, it is the strongly positive Q contributions of the plunge through images in SBS3, which appear for a very short time interval, that allows one to observe the smooth transitions previously mentioned in the EVPA. 

An additional distinguishing factor between SBS1, SBS2, and Schwarzschild BH stands on the range of the values of U. While both SBS1 and SBS2 models present similar negative intervals of Q, the QU-plane shows that SBS1 presents larger negative values of U, which correspond to the slow transition in the EVPA from $t\sim 3T$ to $t\sim 0.7T$, whereas SBS2 presents larger positive values of U, corresponding to the slow transition in the EVPA from $t\sim 0.4T$ to $t\sim 0.6T$. In contrast, for the Schwarzschild BH, the interval of U is more symmetric, not reaching values as large as those observed in SBS1 and SBS2.

Our results thus indicate that inclination plays a crucial role in the differentiation between the models analyzed. Indeed, whereas for low inclination the most compact SBS3 configuration seems to be the one that most closely reproduces the polarimetric observables of a Schwarzschild BH, the sharp deviations on the temporal EVPA and QU-loops induced by the positive Q polarization of the additional plunge-through images of the high inclination configuration cause its polarimetric observables to prominently deviate from their Schwarzschild counterparts. Either way, these results show that the analysis of polarimetric observables allows for the distinction between models for compact objects that are potentially indistinguishable from an optical unpolarized point of view, thus providing an additional tool to assess the validity of these alternative models and to constrain them through a comparison with experimental data.

\section{Conclusions}\label{sec:concl}

In this work we have analyzed the polarimetric observables, namely the time-integrated Stokes parameters I, Q and U, the temporal EVPA, and the temporal QU-loops, of a spherical light source emitting synchrotron radiation while orbiting central solitonic boson star configuration. We have considered three boson stars with different compacticities, one that is dilute (SBS1), one that is close to ultra-compact but still without critical curves (SBS2), and one that is ultra-compact (SBS3). Our results indicate that the polarimetric observables are highly dependent on the compacticity of the central compact object.

The analysis of the integrated Stokes parameters indicates that the polarimetric properties of the primary images are quite independent of the model used for the central object, regardless of the latter being a BH or a boson star. This indicates that it is the additional structure of higher-order images, namely the secondary and plunge-through images, that induces differences in these observables that allows one to distinguish between different models. Indeed, the contribution of the plunge-through images at low inclination can change the fraction of polarization along the orbit in comparison with the Schwarzschild BH. Interestingly, if higher-order images are absent, as it happens for the SBS1 model, the two observed QU-loops for low inclination observations have roughly the same size, which stands in contradiction with recent observations from ALMA \cite{Wielgus2022} and GRAVITY\cite{GRAVITY:2023avo}. We consequently conclude that dilute models incapable of producing secondary images at low inclinations are effectively ruled out by these observations.

While the differences observed in the temporal EVPA and QU-loops at low inclinations are mostly quantitative (with the notable exception of the QU-loops for the SBS1 model mentioned in the previous paragraph), this work shows that inclination plays a crucial role in distinguishing between the different models. Indeed, for high inclination observations, the polarimetric observables present large qualitative differences between the models considered, including the number and time interval between EVPA inversions, and the shape and size of the QU-loops. It can thus be anticipated that, while the current low-inclination observations may only exclude dilute models and remain incapable of constraining more compact models, future higher-inclination observations could more effectively impose strong constrains on these models and affirm which are more likely explicative of the nature of the observed supermassive compact objects.





\begin{thebibliography}{99}

\bibitem{LIGOScientific:2016aoc}
B.~P.~Abbott \textit{et al.} [LIGO Scientific and Virgo],
Phys. Rev. Lett. \textbf{116}, no.6, 061102 (2016)
doi:10.1103/PhysRevLett.116.061102
[arXiv:1602.03837 [gr-qc]].

\bibitem{LIGOScientific:2021djp}
R.~Abbott \textit{et al.} [LIGO Scientific, VIRGO and KAGRA],
[arXiv:2111.03606 [gr-qc]].

\bibitem{KAGRA:2021vkt}
R.~Abbott \textit{et al.} [KAGRA, VIRGO and LIGO Scientific],
Phys. Rev. X \textbf{13}, no.4, 041039 (2023)
doi:10.1103/PhysRevX.13.041039
[arXiv:2111.03606 [gr-qc]].

\bibitem{EventHorizonTelescope:2019dse}
K.~Akiyama \textit{et al.} [Event Horizon Telescope],
Astrophys. J. Lett. \textbf{875}, L1 (2019)
doi:10.3847/2041-8213/ab0ec7
[arXiv:1906.11238 [astro-ph.GA]].

\bibitem{EventHorizonTelescope:2021bee}
K.~Akiyama \textit{et al.} [Event Horizon Telescope],
Astrophys. J. Lett. \textbf{910}, no.1, L12 (2021)
doi:10.3847/2041-8213/abe71d
[arXiv:2105.01169 [astro-ph.HE]].

\bibitem{EventHorizonTelescope:2022wkp}
K.~Akiyama \textit{et al.} [Event Horizon Telescope],
Astrophys. J. Lett. \textbf{930}, no.2, L12 (2022)
doi:10.3847/2041-8213/ac6674

\bibitem{GRAVITY:2020lpa}
M.~Baub\"ock \textit{et al.} [GRAVITY],
Astron. Astrophys. \textbf{635}, A143 (2020)
doi:10.1051/0004-6361/201937233
[arXiv:2002.08374 [astro-ph.HE]].

\bibitem{GRAVITY:2023avo}
R.~Abuter \textit{et al.} [GRAVITY],
Astron. Astrophys. \textbf{677} (2023) L10.

\bibitem{Will:2014kxa}
C.~M.~Will,
Living Rev. Rel. \textbf{17}, 4 (2014)
doi:10.12942/lrr-2014-4
[arXiv:1403.7377 [gr-qc]].

\bibitem{Yagi:2016jml}
K.~Yagi and L.~C.~Stein,
Class. Quant. Grav. \textbf{33}, no.5, 054001 (2016)
doi:10.1088/0264-9381/33/5/054001
[arXiv:1602.02413 [gr-qc]].

\bibitem{Kerr:1963ud}
R.~P.~Kerr,
Phys. Rev. Lett. \textbf{11}, 237-238 (1963)
doi:10.1103/PhysRevLett.11.237

\bibitem{Penrose:1964wq}
R.~Penrose,
Phys. Rev. Lett. \textbf{14}, 57-59 (1965)
doi:10.1103/PhysRevLett.14.57

\bibitem{Penrose:1969pc}
R.~Penrose,
Riv. Nuovo Cim. \textbf{1}, 252-276 (1969)
doi:10.1023/A:1016578408204

\bibitem{Romero:2013ag}
G.~E.~Romero,
Foundations of science \textbf{18}, 297 (2013)
[arXiv:1210.2427 [physics.gen-ph]].

\bibitem{Hawking:1976ra}
S.~W.~Hawking,
Phys. Rev. D \textbf{14}, 2460-2473 (1976)
doi:10.1103/PhysRevD.14.2460

\bibitem{Cardoso:2019rvt}
V.~Cardoso and P.~Pani,
Living Rev. Rel. \textbf{22}, no.1, 4 (2019)
doi:10.1007/s41114-019-0020-4
[arXiv:1904.05363 [gr-qc]].

\bibitem{Visinelli:2021uve}
L.~Visinelli,
Int. J. Mod. Phys. D \textbf{30}, no.15, 2130006 (2021)
doi:10.1142/S0218271821300068
[arXiv:2109.05481 [gr-qc]].

\bibitem{Liebling:2012fv}
S.~L.~Liebling and C.~Palenzuela,
Living Rev. Rel. \textbf{26}, no.1, 1 (2023)
doi:10.1007/s41114-023-00043-4
[arXiv:1202.5809 [gr-qc]].

\bibitem{Brito:2015yga}
R.~Brito, V.~Cardoso and H.~Okawa,
Phys. Rev. Lett. \textbf{115}, no.11, 111301 (2015)
doi:10.1103/PhysRevLett.115.111301
[arXiv:1508.04773 [gr-qc]].

\bibitem{Brito:2015yfh}
R.~Brito, V.~Cardoso, C.~F.~B.~Macedo, H.~Okawa and C.~Palenzuela,
Phys. Rev. D \textbf{93}, no.4, 044045 (2016)
doi:10.1103/PhysRevD.93.044045
[arXiv:1512.00466 [astro-ph.SR]].

\bibitem{Khlopov:1985fch}
M.~Y.~Khlopov, B.~A.~Malomed, I.~B.~Zeldovich and Y.~B.~Zeldovich,
Mon. Not. Roy. Astron. Soc. \textbf{215}, no.4, 575-589 (1985)
doi:10.1093/mnras/215.4.575

\bibitem{Brito:2015pxa}
R.~Brito, V.~Cardoso, C.~A.~R.~Herdeiro and E.~Radu,
Phys. Lett. B \textbf{752}, 291-295 (2016)
doi:10.1016/j.physletb.2015.11.051
[arXiv:1508.05395 [gr-qc]].

\bibitem{Cardoso:2021ehg}
V.~Cardoso, C.~F.~B.~Macedo, K.~i.~Maeda and H.~Okawa,
Class. Quant. Grav. \textbf{39}, no.3, 034001 (2022)
doi:10.1088/1361-6382/ac41e7
[arXiv:2112.05750 [gr-qc]].

\bibitem{Cao:2016zbh}
Z.~Cao, A.~Cardenas-Avendano, M.~Zhou, C.~Bambi, C.~A.~R.~Herdeiro and E.~Radu,
JCAP \textbf{10}, 003 (2016)
doi:10.1088/1475-7516/2016/10/003
[arXiv:1609.00901 [gr-qc]].

\bibitem{Dove:1997ei}
J.~B.~Dove, J.~Wilms, M.~Maisack and M.~C.~Begelman,
Astrophys. J. \textbf{487}, 759 (1997)
doi:10.1086/304647
[arXiv:astro-ph/9705130 [astro-ph]].


\bibitem{Hui:2016ltb}
L.~Hui, J.~P.~Ostriker, S.~Tremaine and E.~Witten,
Phys. Rev. D \textbf{95}, no.4, 043541 (2017)
doi:10.1103/PhysRevD.95.043541
[arXiv:1610.08297 [astro-ph.CO]].

\bibitem{Palenzuela:2017kcg}
C.~Palenzuela, P.~Pani, M.~Bezares, V.~Cardoso, L.~Lehner and S.~Liebling,
Phys. Rev. D \textbf{96}, no.10, 104058 (2017)
doi:10.1103/PhysRevD.96.104058
[arXiv:1710.09432 [gr-qc]].

\bibitem{Cardoso:2017cqb}
V.~Cardoso and P.~Pani,
Nature Astron. \textbf{1}, no.9, 586-591 (2017)
doi:10.1038/s41550-017-0225-y
[arXiv:1709.01525 [gr-qc]].

\bibitem{Cardoso:2016oxy}
V.~Cardoso, S.~Hopper, C.~F.~B.~Macedo, C.~Palenzuela and P.~Pani,
Phys. Rev. D \textbf{94}, no.8, 084031 (2016)
doi:10.1103/PhysRevD.94.084031
[arXiv:1608.08637 [gr-qc]].

\bibitem{Cardoso:2016rao}
V.~Cardoso, E.~Franzin and P.~Pani,
Phys. Rev. Lett. \textbf{116}, no.17, 171101 (2016)
[erratum: Phys. Rev. Lett. \textbf{117}, no.8, 089902 (2016)]
doi:10.1103/PhysRevLett.116.171101
[arXiv:1602.07309 [gr-qc]].

\bibitem{Postnikov:2010yn}
S.~Postnikov, M.~Prakash and J.~M.~Lattimer,
Phys. Rev. D \textbf{82}, 024016 (2010)
doi:10.1103/PhysRevD.82.024016
[arXiv:1004.5098 [astro-ph.SR]].

\bibitem{Cardoso:2017cfl}
V.~Cardoso, E.~Franzin, A.~Maselli, P.~Pani and G.~Raposo,
Phys. Rev. D \textbf{95}, no.8, 084014 (2017)
doi:10.1103/PhysRevD.95.084014
[arXiv:1701.01116 [gr-qc]].

\bibitem{Herdeiro:2021lwl}
C.~A.~R.~Herdeiro, A.~M.~Pombo, E.~Radu, P.~Cunha, V.P. and N.~Sanchis-Gual,
JCAP \textbf{04}, 051 (2021)
doi:10.1088/1475-7516/2021/04/051
[arXiv:2102.01703 [gr-qc]].

\bibitem{Rosa:2022tfv}
J.~L.~Rosa and D.~Rubiera-Garcia,
Phys. Rev. D \textbf{106}, no.8, 084004 (2022)
doi:10.1103/PhysRevD.106.084004
[arXiv:2204.12949 [gr-qc]].

\bibitem{Rosa:2022toh}
J.~L.~Rosa, P.~Garcia, F.~H.~Vincent and V.~Cardoso,
Phys. Rev. D \textbf{106}, no.4, 044031 (2022)
doi:10.1103/PhysRevD.106.044031
[arXiv:2205.11541 [gr-qc]].

\bibitem{Rosa:2023qcv}
J.~L.~Rosa, C.~F.~B.~Macedo and D.~Rubiera-Garcia,
Phys. Rev. D \textbf{108}, no.4, 044021 (2023)
doi:10.1103/PhysRevD.108.044021
[arXiv:2303.17296 [gr-qc]].

\bibitem{Olivares:2018abq}
H.~Olivares, Z.~Younsi, C.~M.~Fromm, M.~De Laurentis, O.~Porth, Y.~Mizuno, H.~Falcke, M.~Kramer and L.~Rezzolla,
Mon. Not. Roy. Astron. Soc. \textbf{497}, no.1, 521-535 (2020)
doi:10.1093/mnras/staa1878
[arXiv:1809.08682 [gr-qc]].

\bibitem{Gjorgjieski:2023qpv}
K.~Gjorgjieski, J.~Kunz, M.~C.~Teodoro, L.~G.~Collodel and P.~Nedkova,
Phys. Rev. D \textbf{107}, no.10, 103043 (2023)
doi:10.1103/PhysRevD.107.103043
[arXiv:2301.00449 [gr-qc]].

\bibitem{Rosa:2024eva}
J.~L.~Rosa, J.~Pelle and D.~P\'erez,
[arXiv:2403.11540 [gr-qc]].

\bibitem{EHT_2021}EHT Collaboration,
The Astrophysical Journal Letters \textbf{910}, L12 (2021,3)
doi:10.3847/2041-8213/abe71d

\bibitem{EHT_2024}EHT Collaboration,
The Astrophysical Journal Letters \textbf{964}, L25 (2024,3)
doi:10.3847/2041-8213/ad2df0

\bibitem{gravity2018}GRAVITY Collaboration, 
Astronomy and Astrophysics \textbf{618} pp. eL10 (2018,10)

\bibitem{gravity2023}Gravity Collaboration, 
Astronomy and Astrophysics \textbf{677} pp. eL10 (2023,9)

\bibitem{EHT2024b}EHT Collaboration,
The Astrophysical Journal Letters. \textbf{964}, L26 (2024,3)
doi:10.3847/2041-8213/ad2df1

\bibitem{Wielgus2022}Wielgus, M., Moscibrodzka, M., Vos, J., Gelles, Z., Mart\'i-Vidal, I., Farah, J., Marchili, N., Goddi, C. \& Messias, H. 
Astronomy and Astrophysics \textbf{665} pp. eL6 (2022,9)

\bibitem{Vincent:2023sbw}
F.~H.~Vincent, M.~Wielgus, N.~Aimar, T.~Paumard and G.~Perrin,
Astron. Astrophys. \textbf{684} (2024), A194
doi:10.1051/0004-6361/202348016
[arXiv:2309.10053 [astro-ph.HE]].

\bibitem{Himwich2020}Himwich, E., Johnson, M., Lupsasca, A. \& Strominger, A. 
Physical Review D. \textbf{101}, e084020 (2020,4)

\bibitem{Palumbo2023}Palumbo, D., Wong, G., Chael, A. \& Johnson, M. 
The Astrophysical Journal Letters \textbf{952}, eL31 (2023,8)

\bibitem{Vincent:2011wz}
F.~H.~Vincent, T.~Paumard, E.~Gourgoulhon and G.~Perrin,
Class. Quant. Grav. \textbf{28} (2011), 225011
doi:10.1088/0264-9381/28/22/225011
[arXiv:1109.4769 [gr-qc]].

\bibitem{Aimar:2023vcs}
N.~Aimar, T.~Paumard, F.~H.~Vincent, E.~Gourgoulhon and G.~Perrin,
Class. Quant. Grav. \textbf{41} (2024) no.9, 095010
doi:10.1088/1361-6382/ad351d
[arXiv:2311.18802 [astro-ph.HE]].

\bibitem{Macedo:2013jja}
C.~F.~B.~Macedo, P.~Pani, V.~Cardoso and L.~C.~B.~Crispino,
Phys. Rev. D \textbf{88} (2013) no.6, 064046
doi:10.1103/PhysRevD.88.064046
[arXiv:1307.4812 [gr-qc]].

\bibitem{Lee:1986ts}
T.~D.~Lee,
Phys. Rev. D \textbf{35} (1987), 3637
doi:10.1103/PhysRevD.35.3637

\bibitem{Lee:1991ax}
T.~D.~Lee and Y.~Pang,
Phys. Rept. \textbf{221} (1992), 251-350
doi:10.1016/0370-1573(92)90064-7

\bibitem{RybickiLightman:1979}
{Rybicki}, George B. and {Lightman}, Alan P.,
1979.

\bibitem{Marszewski:2021}
Marszewski, A., Prather, B., Joshi, A., Pandya, A. \& Gammie, C.,
The Astrophysical Journal \textbf{921}, e17 (2021,11)

\bibitem{Gelles:2021}
Gelles, Z., Himwich, E., Johnson, M. \& Palumbo, D. 
Physical Review D \textbf{104}, e044060 (2021,8)


\end{thebibliography}
\end{document}